\documentclass[%
 reprint,
 aps,
 superscriptaddress,
 raggedbottom,
 prl]{revtex4-2}
\usepackage{CJK}

\usepackage{amsmath,amssymb,mathtools,extarrows,mathrsfs}
\usepackage{tikz,float,subcaption,graphbox}

\usepackage{titlesec}
\titleformat{\section}{\centering\normalsize\normalfont\bf}{\thesection}{0em}{}
\usepackage{hyperref}

\begin{document}

\begin{CJK*}{UTF8}{}
\CJKfamily{gbsn}
\title{Constructibility of AdS supergluon amplitudes}

\author{Qu Cao (曹趣)}
\email{qucao@zju.edu.cn}
\affiliation{CAS Key Laboratory of Theoretical Physics, Institute of Theoretical Physics, Chinese Academy of Sciences, Beijing 100190, China}
\affiliation{Zhejiang Institute of Modern Physics, Department of Physics, Zhejiang University, Hangzhou, Zhejiang 310027, China}
\author{Song He (何颂)}
\email{songhe@itp.ac.cn}
\affiliation{CAS Key Laboratory of Theoretical Physics, Institute of Theoretical Physics, Chinese Academy of Sciences, Beijing 100190, China}
\affiliation{School of Fundamental Physics and Mathematical Sciences, Hangzhou Institute for Advanced Study, Hangzhou, Zhejiang 310024, China}
\affiliation{International Centre for Theoretical Physics Asia-Pacific, Beijing 100190, China}
\affiliation{Peng Huanwu Center for Fundamental Theory, Hefei, Anhui 230026, China}
\author{Yichao Tang (唐一朝)}
\email{tangyichao@itp.ac.cn}
\affiliation{CAS Key Laboratory of Theoretical Physics, Institute of Theoretical Physics, Chinese Academy of Sciences, Beijing 100190, China}
\affiliation{School of Physical Sciences, University of Chinese Academy of Sciences, No.19A Yuquan Road, Beijing 100049, China}

\begin{abstract}
We prove that all tree-level $n$-point supergluon (scalar) amplitudes in AdS$_5$ can be recursively constructed, using factorization and flat-space limit. Our method is greatly facilitated by a natural R-symmetry basis for planar color-ordered amplitudes, which reduces the latter to ``partial amplitudes" with simpler pole structures and factorization properties. Given the $n$-point scalar amplitude, we first extract spinning amplitudes with $n{-}2$ scalars and one gluon by imposing ``gauge invariance", and then use a special ``no-gluon kinematics" to determine the $(n{+}1)$-point scalar amplitude completely (which in turn contains the $n$-point single-gluon amplitude). Explicit results of up to 8-point scalar amplitudes and up to 6-point single-gluon amplitudes are included as Supplemental Material.
\end{abstract}

\maketitle
\end{CJK*}

\section{Introduction}
Recent years have witnessed remarkable progress in computing and revealing new structures of holographic correlators, or ``scattering amplitudes'' in AdS space, at both tree~\cite{Rastelli:2016nze, Rastelli:2017udc,Rastelli:2017ymc,Zhou:2017zaw,Rastelli:2019gtj,Goncalves:2019znr,Alday:2020lbp,Alday:2020dtb,Zhou:2021gnu,Goncalves:2023oyx} and loop~\cite{Alday:2017xua,Aprile:2017bgs,Aprile:2017qoy,Aprile:2019rep,Alday:2019nin,Huang:2021xws,Drummond:2022dxw} level. Although more focus has been on supergravity amplitudes in AdS, explicit results have also been obtained for ``supergluon" tree amplitudes up to $n=6$~\cite{Zhou:2018ofp,Alday:2021odx,Alday:2022lkk,Bissi:2022wuh,Alday:2023kfm} in AdS super-Yang-Mills (sYM) theories (see~\cite{Alday:2021ajh,Huang:2023oxf,Huang:2023ppy} for loop-level results). In this Letter, we ask the interesting question about the ``constructibility" of higher-point supergluon amplitudes purely from lower-point ones, and along the way we reveal nice structures for these amplitudes to all $n$.

The natural language for holographic correlators is the Mellin representation~\cite{Mack:2009mi,Penedones:2010ue,Fitzpatrick:2011ia}. Mellin tree amplitudes are rational functions of Mellin variables. They can be determined by the residues at all physical poles (and pole at infinity encoded in the flat-space limit~\cite{Alday:2023kfm}), which for sYM are given by factorization with scalar and gluon exchanges~\cite{Goncalves:2014rfa}. These allowed the authors of \cite{Alday:2022lkk,Alday:2023kfm} to bootstrap the supergluon amplitudes up to six point.

However, naively using factorization to bootstrap higher-point supergluon amplitudes is difficult, because we lack data of higher-point amplitudes involving spinning particles, which are needed to compute gluon-exchange contributions. We overcome this problem by getting ``more'' out of scalar-exchange contributions.

\begin{figure}[H]
    \centering
    \includegraphics[scale=0.65]{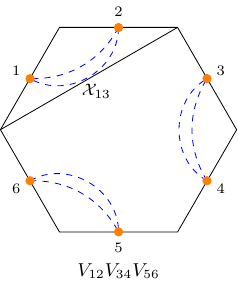}\hspace{1em}\includegraphics[scale=0.65]{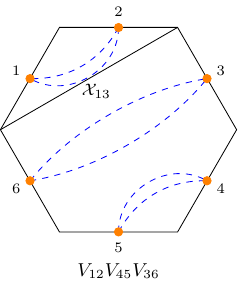}
    
    \includegraphics[scale=0.65]{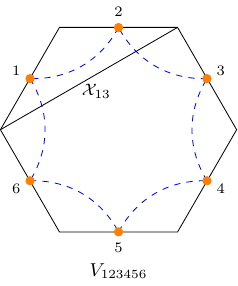}\hspace{1em}\includegraphics[scale=0.65]{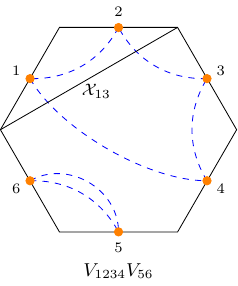}\hspace{1em}\includegraphics[scale=0.65]{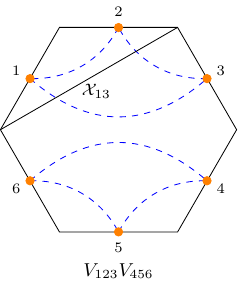}
    \caption{$\mathcal M_6^{(s)}$ R-structures compatible (above) and incompatible (below) with $\mathcal X_{13}$.}
    \label{fig:compatR}
\end{figure}

On one hand, we recognize a natural R-symmetry basis (Fig.~\ref{fig:compatR}) built from SU$(2)_R$ traces compatible with color ordering. Knowing lower-point scalar amplitudes, we are able to isolate the gluon-exchange contributions in factorization channels compatible with the trace structure. This enables us to extract the $(n-1)$-point single-gluon amplitude from the $n$-point scalar amplitude.

On the other hand, we identify certain ``no-gluon kinematics'' which is a consequence of the ``gauge invariance'' of single-gluon amplitudes. Regardless of the precise form of single-gluon amplitudes, at these special kinematic points, gluon exchanges are forbidden, imposing a powerful constraint on the amplitude.

Combining these two realizations, we devise a recursive algorithm~\eqref{eq:sequence} to obtain all-multiplicity supergluon tree amplitudes: start from the $n$-point scalar amplitude, extract from it the $(n-1)$-point single-gluon amplitude, and use these (sufficient) information to construct the $(n+1)$-point scalar amplitude. We include explicit results of up to 8-point scalar amplitudes and up to 6-point single-gluon amplitudes in the Supplemental Material~\footnote{See Supplemental Material at \href{http://link.aps.org/supplemental/10.1103/PhysRevLett.133.021605}{http://link.aps.org/ supplemental/10.1103/PhysRevLett.133.021605} for details of the Mellin factorization formula, explicit results for up to $n=6$ amplitudes, as well as a .txt file containing explicit results of $\mathcal M_{\leq8}^{(s)}$ and $\mathcal M_{\leq6}^{(v)}$.}.

\section{Organization of Mellin amplitudes}
We are interested in the $n$-point supergluon amplitudes in ${\rm AdS}_5/{\rm CFT}_4$, which arise as the low energy description of many different theories~\cite{Fayyazuddin:1998fb,Aharony:1998xz,Karch:2002sh,Alday:2021odx}. For concreteness, consider the D3-D7-brane system in Type IIB string theory in the probe limit (number $N_f$ of D7-branes much less than number $N_c$ of D3-branes)~\cite{Karch:2002sh}. On the world volume of D3-branes, we have an $\mathcal N=2$ SCFT, while on the world volume of D7-branes, gravity decouples at tree level and we have $\mathcal N=1$ sYM on ${\rm AdS}_5\times S^3$~\footnote{The gravitational coupling is proportional to $1/N_c$, which is much smaller than the (super)gluon self coupling proportional to $1/\sqrt{N_c}$. Hence, gravity decouples at tree level, i.e., leading $1/N_c$ order.}. The system has a symmetry $G_F={\rm SU}(N_f)$~\footnote{Other theories such as those arising in F-theories~\cite{Fayyazuddin:1998fb,Aharony:1998xz} lead to different $G_F$, but otherwise the effective descriptions are the same.} which is global on the boundary and local in the bulk.

We study the connected correlator of half-BPS operators $\mathcal O^a(x,v)$ with dimension $\Delta=2$:
\begin{gather}
    G_n^{(s)a_1\cdots a_n}=\langle\mathcal O^{a_1}(x_1,v_1)\cdots\mathcal O^{a_n}(x_n,v_n)\rangle,\\
    \mathcal O^a(x,v)=\mathcal O^{a;\alpha_1\alpha_2}(x)v^{\beta_1}v^{\beta_2}\epsilon_{\alpha_1\beta_1}\epsilon_{\alpha_2\beta_2}.
\end{gather}
Here, $a_i=1,\cdots,\dim G_F$ are adjoint indices of $G_F$, and $v^\beta$ ($\alpha_i,\beta_i=1,2$) are auxiliary SU$(2)_R$-spinors which extracts the R-spin-1 part of $\mathcal O^{a;\alpha_1\alpha_2}(x)$. The superscript $^{(s)}$ reminds us that $G_n^{(s)}$ is a correlator of scalar operators. For convenience, we also introduce the single-gluon correlators $G_n^{(v)}$ involving the Noether current $\mathcal J_\mu^a(x)$ of $G_F$, an SU$(2)_R$-singlet with dimension $\Delta=3$:
\begin{equation}
    G_{n;\mu}^{(v)a_1\cdots a_n}=\langle\mathcal O^{a_1}(x_1)\cdots\mathcal O^{a_{n-1}}(x_{n-1})\mathcal J_\mu^{a_n}(x_n)\rangle.
\end{equation}
The bulk dual of $\mathcal O^a$ is $\phi_m^a$ for $m=1,2,3$ (``supergluon''), and the bulk dual of $\mathcal J_\mu^a$ is $A_\mu^a$ (``gluon''). Together, they compose the lowest Kaluza-Klein mode of the $G_F$ gauge field on ${\rm AdS}_5\times S^3$. It can be shown that these are all the fields needed for $G_n^{(s)}$ at tree level~\footnote{Within the supermultiplet containing $\mathcal O^a$ and $\mathcal J_\mu^a$, all other primaries are charged under U$(1)_r$, the Abelian part of the $\mathcal N=2$ R-symmetry. Other half-BPS supermultiplets are dual to higher Kaluza-Klein modes, and they are charged under ${\rm SU}(2)_L$ which is part of the isometry group ${\rm SO}(4)={\rm SU}(2)_L\times{\rm SU}(4)_R$ of $S^3$. The operators $\mathcal O^a$ and $\mathcal J_\mu^a$ are special in that they are neutral under U$(1)_r$ and SU$(2)_L$. Hence, all other fields can only appear in pairs and contribute at loop level.}.

The color decomposition for tree amplitudes in AdS space is identical to that for flat-space amplitudes~\cite{DelDuca:1999rs}: we have color-ordered amplitudes as coefficients in front of traces of generators $T^a$ in the adjoint representation:
\begin{equation}
    G_n^{a_1\cdots a_n}=\sum_{\mathclap{\sigma\in S_{n-1}}}{\rm tr}(T^{a_1}T^{a_2^{\sigma}}\cdots T^{a_{n-1}^{\sigma}}T^{a_n^{\sigma}})G_{1\sigma}\,,
\end{equation}
where ${\sigma}$ denotes a permutation of $\{2,\cdots, n\}$. Cyclic and reflection symmetry of the traces implies
\begin{equation}\label{eq:dihe}
    G_{12\cdots n}=G_{2\cdots n1}=(-)^nG_{n\cdots21}\,.
\end{equation}
We will focus on $G_{12\cdots n}$ since any color-ordered amplitude can then be obtained by relabeling.

The natural language to describe such CFT correlators is the Mellin representation~\cite{Mack:2009mi}. For scalar amplitudes,
\begin{equation}
    G_{12\cdots n}^{(s)}=\int[{\rm d}\delta]\mathcal M_n^{(s)}(\{\delta_{ij}\},\{v_i\})\prod_{i<j}\frac{\Gamma(\delta_{ij})}{(-2P_i\cdot P_j)^{\delta_{ij}}},
\end{equation}
and for single-gluon amplitudes~\cite{Goncalves:2014rfa}:
\begin{gather}
    G_{12\cdots n}^{(v)}=\int[{\rm d}\delta]\sum_{\ell=1}^{n-1}(Z_n\cdot P_\ell)\mathcal M_n^{(v)\ell}\prod_{i<j}\frac{\Gamma(\delta_{ij}+\delta_i^\ell\delta_j^n)}{(-2P_i\cdot P_j)^{\delta_{ij}+\delta_i^\ell\delta_j^n}},\\
    \text{where }\sum_{\ell=1}^{n-1}\delta_{\ell n}\mathcal M_n^{(v)\ell}=0.\label{eq:gaugeinv}
\end{gather}
Note that here $\delta_{i}^{l}$ is the Kronecker delta. We have used the embedding formalism following~\cite{Goncalves:2014rfa}, where $P_i\cdot P_j=-\frac12(x_i-x_j)^2$ and $Z_n\cdot P_\ell$ encodes the Lorentz tensor structure of $\mathcal J_\mu^a$. The Mellin variables are constrained as if $\delta_{ij}=p_i\cdot p_j$ for auxiliary momenta satisfying $\sum_ip_i=0$ and $p_i^2=-\tau_i=-2$, with conformal twist $\tau_i:=\Delta_i-J_i$ ($J$ is the spin of an operator). Since $\mathcal J$ and $\mathcal O$ have the same twist, they are described by the same ``kinematics''.

\begin{figure}[H]
    \vspace{-1em}
    \centering
    \includegraphics[scale=0.7]{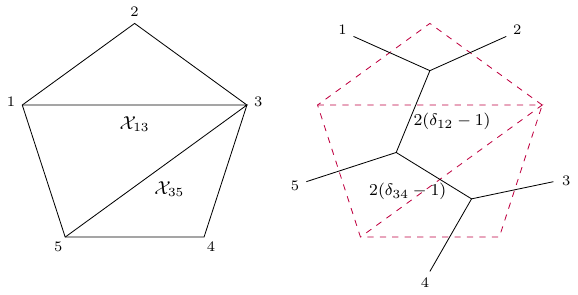}
    \vspace{-1em}
    \caption{Planar variables and dual skeleton graph for $n=5$.}
    \label{fig:n-gonXij}
\end{figure}

Only the $\frac12n(n-3)$ $\delta_{ij}$'s are independent. Inspired by flat space~\cite{Arkani-Hamed:2017mur}, it proves convenient to introduce $\frac12n(n-3)$ planar variables (with $\delta_{ii}\equiv-2$)
\begin{equation}
{\cal X}_{ij}:=2+\sum_{i\leq k, l<j} \delta_{kl}=2+\Bigg(\sum_{i\leq k<j}p_k\Bigg)^2,
\end{equation}
where we have ${\cal X}_{i,j}={\cal X}_{j,i}$ with special cases ${\cal X}_{i,i{+}1}=0$ and ${\cal X}_{i,i}\equiv 2$. The inverse transform which motivated the associahedron in~\cite{Arkani-Hamed:2017mur, Arkani-Hamed:2019vag} reads:
\begin{equation}
    -2\delta_{ij}=\mathcal X_{i,j}+\mathcal X_{i+1,j+1}-\mathcal X_{i,j+1}-\mathcal X_{i+1,j}.
\end{equation}
Planar variables correspond to $n$-gon chords (Figure~\ref{fig:n-gonXij}).

The planar variables are particularly suited for factorization~\cite{Goncalves:2014rfa} of color-ordered amplitudes. Since all relevant fields have $\tau=2$, schematically,
\begin{equation}\label{eq:facschem}
    \mathcal M_{12\cdots n}\sim\frac{\mathcal M_{1\cdots(k-1)I}^{(m)}\mathcal M_{k\cdots nI}^{(m)}}{-(\mathcal X_{1k}+2m)},\quad m=0,1,2,\cdots
\end{equation}
where a pole at $\mathcal X_{1k}=-2m$ corresponds to the exchange of a level-$m$ descendant. By induction, all simultaneous poles of $\mathcal M_n$ consist of \emph{compatible} planar variables (non-intersecting chords), which gives a (partial) triangulation of the $n$-gon dual to  planar skeleton graphs (Figure~\ref{fig:n-gonXij}).

\begin{figure}[H]
    \centering
    \includegraphics[scale=0.65]{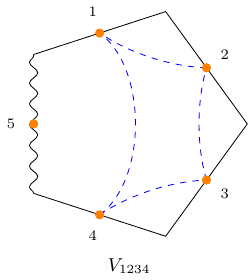}\hspace{1ex}\includegraphics[scale=0.65]{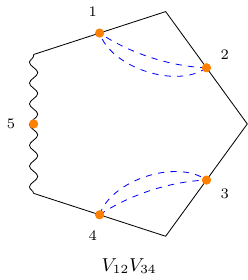}\hspace{1ex}\includegraphics[scale=0.65]{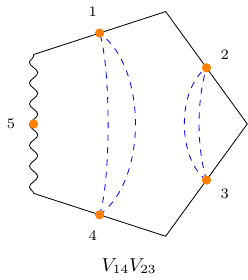}
    \caption{$\mathcal M_5^{(v)}$ R-structures.}
    \label{fig:gluonR}
\end{figure}

Another advantage of working with color-ordered amplitude is a natural basis for the R-charge structures. Let us define SU$(2)_R$ trace as $V_{i_1i_2\cdots i_r}:=\langle i_1i_2\rangle\langle i_2i_3\rangle\cdots\langle i_ri_1\rangle$ where $\langle ij\rangle:=v_i^\alpha v_j^\beta\epsilon_{\alpha\beta}$. The Schouten identity $\langle ik\rangle\langle jl\rangle=\langle ij\rangle\langle kl\rangle+\langle il\rangle\langle jk\rangle$ enables us to expand any R-structure to products of non-crossing cycles or SU$(2)_R$ traces:
\begin{align}
{\cal M}_n^{(s)}&=\sum_{\substack{\text{non-crossing}\\\text{partition }\pi\\\text{of }\{1,\cdots,n\}}}\Bigg(\prod_{\text{cycle }\tau\,\in\,\pi}V_\tau\Bigg)M^{(s)}_n(\pi),\\
{\cal M}_n^{(v)\ell}&=\sum_{\substack{\text{non-crossing}\\\text{partition }\pi\\\text{of }\{1,\cdots,n{-}1\}}}\Bigg(\prod_{\text{cycle }\tau\,\in\,\pi}V_\tau\Bigg)M^{(v)\ell}_n(\pi).
\end{align}
For example, (Figure~\ref{fig:gluonR})
\begin{align*}
    \mathcal M_4^{(s)}&=M_4^{(s)}(1234)V_{1234}\nonumber\\
    &+M_4^{(s)}(12;34)V_{12}V_{34}+M_4^{(s)}(14;23)V_{14}V_{23},\\
    \mathcal M_4^{(v)\ell}&=M_4^{(v)\ell}(123)V_{123},\\
    \mathcal M_5^{(s)}&=M_5^{(s)}(12345)V_{12345}\nonumber\\
    &+M_5^{(s)}(12;345)V_{12}V_{345}+\text{cyclic},\\
    \mathcal M_5^{(v)\ell}&=M_5^{(v)\ell}(1234)V_{1234}\nonumber\\
    &+M_5^{(v)\ell}(12;34)V_{12}V_{34}+M_5^{(v)\ell}(14;23)V_{14}V_{23}.
\end{align*}
Because a length-$L$ trace picks up $(-)^L$ under reflection, for scalar amplitudes this cancels the sign in~\eqref{eq:dihe} while for single-gluon amplitudes the net result is a minus sign:
\begin{align*}
    M_4^{(s)}(12;34)&\xlongequal{\text{ref}}M_4^{(s)}(21;43)\xlongequal{\text{cyc}}M_4^{(s)}(14;23),\\
    M_5^{(v)}(12;34)&\xlongequal{\text{ref}}-M_5^{(v)}(21;43),\\
    M_5^{(v)}(12;34)&\text{ unrelated to }M_5^{(v)}(14;23).
\end{align*}

For scalar amplitudes with $n=6,7$, we additionally have triple-trace R-structures, and for $n\geq8$ we need quadruple-trace R-structures. The number of linearly independent R-structures for $\mathcal M_n^{(s)}$ or $\mathcal M_{n+1}^{(v)}$ is $r_n=1,3,6,15,36,91,\cdots$ (Riordan numbers~\footnote{OEIS database: \href{https://oeis.org/A005043}{https://oeis.org/A005043}.}).

\section{Properties of Mellin amplitudes}

\begin{figure}[H]
    \centering
    \includegraphics[scale=0.65]{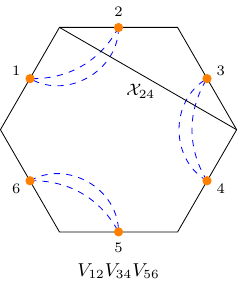}\hspace{1em}\includegraphics[scale=0.65]{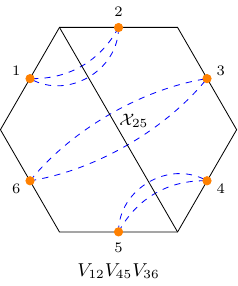}
    \caption{Vanishing R-structures.}
    \label{fig:vanishR}
\end{figure}

\paragraph{Factorization}Different exchanged fields contribute to different R-structures. For a given channel, say $\mathcal X_{1k}$, we distinguish the compatible R-structures $\pi$ (none of the cycles $\tau$ intersect $\mathcal X_{1k}$) from the incompatible ones (Figure~\ref{fig:compatR}). For scalar exchanges, \eqref{eq:facschem} reads
\begin{equation}
    \mathop{\rm Res}_{\mathcal X_{1k}=-2m}^{(s)}\mathcal M_n^{(s)}=\mathcal N_s^{(m)}\texttt{glueR}\left(\mathcal M_{1\cdots(k-1)I}^{(s)(m)}\mathcal M_{k\cdots nI}^{(s)(m)}\right).
\end{equation}
Here, $\mathcal N_s^{(m)}=2$, and $\mathcal M_{1\cdots(k-1)I}^{(s)(m)}$ is a shifted version of the scalar amplitude $\mathcal M_{1\cdots(k-1)I}^{(s)}$:
\begin{equation}\label{eq:shift}
    \mathcal M_{1\cdots(k-1)I}^{(s)(m)}=\sum_{\substack{n_{ab}\geq0\\\sum n_{ab}=m}}\mathcal M_{1\cdots(k-1)I}^{(s)}(\delta_{ab}+n_{ab})\prod_{1\leq a<b<k}\frac{(\delta_{ab})_{n_{ab}}}{n_{ab}!}.
\end{equation}
$\mathcal M_{k\cdots nI}^{(s)(m)}$ is defined similarly. The operation $\texttt{glueR}$ glues together the traces. Note that there is the 1-1 correspondence of R-structures in amplitudes and the operator product expansion (OPE):
\begin{gather*}
    \langle\mathcal O(v_I)\mathcal O\cdots\mathcal O\rangle\supset\text{something}\times V_{ia\cdots bjI}\\
    \Updownarrow\\
    \mathcal O\cdots\mathcal O\supset\text{something}\times\langle ia\rangle\cdots\langle bj\rangle v_i^{(\alpha}v_j^{\beta)}\mathcal O_{\alpha\beta}
\end{gather*}
Since $\langle\mathcal O_{\alpha\beta}\mathcal O_{\gamma\delta}\rangle=\frac12(\epsilon_{\alpha\gamma}\epsilon_{\beta\delta}+\epsilon_{\alpha\delta}\epsilon_{\beta\gamma})$, we have
\begin{align}
v_i^{(\alpha}v_j^{\beta)}v_k^{(\gamma}v_l^{\delta)}\langle\mathcal O_{\alpha\beta}\mathcal O_{\gamma\delta}\rangle
    =\langle il\rangle\langle jk\rangle-\frac12\langle ij\rangle\langle lk\rangle.
\end{align}
which implies the following gluing rule:
\begin{equation}
    \texttt{glueR}:\ V_{i\cdots jI}\otimes V_{Ik\cdots l}\mapsto V_{i\cdots jk\cdots l}-\frac12V_{i\cdots j}V_{k\cdots l}.
\end{equation}
We see that scalar exchanges contribute to both compatible and incompatible R-structures. R-structures with more than one cycle intersecting $\mathcal X_{1k}$ vanish (Figure~\ref{fig:vanishR}).

For gluon exchanges, \eqref{eq:facschem} reads
\begin{equation}\label{eq:gluonfac}
    \mathop{\rm Res}_{\mathcal X_{1k}=-2m}^{(v)}\mathcal M_n^{(s)}=\mathcal N_v^{(m)}\sum_{a=1}^{k-1}\sum_{i=k}^n\delta_{ai}\mathcal M_{1\cdots(k-1)I}^{(v)(m)a}\mathcal M_{k\cdots nI}^{(v)(m)i}.
\end{equation}
Here, $\mathcal N_v^{(m)}=-\frac3{1+m}$, and $\mathcal M_{1\cdots(k-1)I}^{(v)(m)a}$ is $\mathcal M_{1\cdots(k-1)I}^{(v)a}$ shifted according to \eqref{eq:shift}. We no longer need $\texttt{glueR}$ because $\mathcal J$ is R-neutral; gluon exchanges contribute to compatible R-structures only.

An important consequence of ``gauge invariance'' \eqref{eq:gaugeinv} is that, at certain \emph{no-gluon kinematics}, gluon exchanges are forbidden completely. To see this, let us denote $\mathcal M_{1\cdots(k-1)I}^{(v)(m)a}\equiv\mathcal L^{(m)a}$ and $\mathcal M_{k\cdots nI}^{(v)(m)i}\equiv\mathcal R^{(m)i}$, and solve $\mathcal L^{(m)1},\mathcal R^{(m)k}$ using \eqref{eq:gaugeinv}. The double sum in \eqref{eq:gluonfac} becomes
\begin{equation*}
    \sum_{a=2}^{k-1}\sum_{i=k+1}^{n}\left(\delta_{ai}-\frac{\delta_{aI}}{\delta_{1I}}\delta_{1i}-\frac{\delta_{iI}}{\delta_{kI}}\delta_{ak}+\frac{\delta_{aI}\delta_{iI}}{\delta_{1I}\delta_{kI}}\delta_{1k}\right)\mathcal L^{(m)a} \mathcal R^{(m)i}.
\end{equation*}
If all $(k-2)(n-k)$ coefficients vanish on the support of $\mathcal X_{1k}=-2m$, gluon exchanges are forbidden, regardless of the detailed form of $\mathcal L^{(m)}$ and $\mathcal R^{(m)}$. The number of conditions equals the number of chords $\mathcal X_{ai}$ ($2\leq a\leq k-1$ and $k+1\leq i\leq n$) crossing $\mathcal X_{1k}$. Hence, the no-gluon conditions translate to $\mathcal X_{ai}$ taking special values $\mathcal X_{ai}^*$:
\begin{align}
    \mathcal E_{ai}^{(m)}&:=\mathcal X_{ai}-\mathcal X_{ai}^*=0,\\
    X_{ai}^*&=m-1+\frac{\mathcal X_{1a}+\mathcal X_{1i}+\mathcal X_{ak}+\mathcal X_{ik}}2\nonumber\\
    &+\frac{(\mathcal X_{1a}-\mathcal X_{ak})(\mathcal X_{1i}-\mathcal X_{ik})}{4(m+1)}.
\end{align}
Since gluon exchanges are forbidden at no-gluon kinematics, scalar exchanges alone fix the residue up to polynomials of $\mathcal E$'s:
\begin{equation}\label{eq:NGfix}
    \mathop{\rm Res}_{\mathcal X_{1k}=-2m}\mathcal M_n =\left.\mathop{\rm Res}_{\mathcal X_{1k}=-2m}^{(s)}\mathcal M_n\right|_{\mathcal X_{ai}=\mathcal X_{ai}^*}+\text{poly}(\mathcal E_{ai}^{(m)}).
\end{equation}

The special case of \eqref{eq:gluonfac} where $k=n-1$ is particularly important. From the 3-point single-gluon amplitude~\footnote{We can easily compute it as follows. First, the R-structure is unique. Second, there are no free Mellin variables, so the amplitude must be a constant. Third, \eqref{eq:gaugeinv} fixes the relative sign to be $-1$. Lastly, we can fix the overall normalization by comparing to the gluon-exchange contribution to $\mathcal M_4^{(s)}$.}:
\begin{equation}
    \mathcal M_{n-1,n,I}^{(v)(0)n-1}=\frac i{\sqrt6}V_{n-1,n},\ \mathcal M_{n-1,n,I}^{(v)(0)n}=-\frac i{\sqrt6}V_{n-1,n},
\end{equation}
we see that
\begin{equation}
    \mathop{\rm Res}_{\mathcal X_{1,n-1}=0}^{(v)}\mathcal M_n^{(s)}=\frac{-3i}{\sqrt6}V_{n-1,n}\sum_{a=1}^{n-2}(\delta_{a,n-1}-\delta_{a,n})\mathcal M_{n-1}^{(v)a}.
\end{equation}
This is similar to the scaffolding relation in~\cite{Arkani-Hamed:2023jry}. If we write the $\delta$'s in terms of $\mathcal X$'s, one can show that for each $2\leq a\leq n-2$,
\begin{equation}
    \mathcal M_{n-1}^{(v)a}-\mathcal M_{n-1}^{(v)a-1}=\frac\partial{\partial\mathcal X_{an}}\left(\frac{i\sqrt{2/3}}{V_{n-1,n}}\mathop{\rm Res}_{\mathcal X_{1,n-1}=0}^{(v)}\mathcal M_n^{(s)}\right).
\end{equation}
Together with \eqref{eq:gaugeinv}, these $(n-3)+1$ equations completely determine $\{\mathcal M_{n-1}^{(v)a}\}_{a=1}^{n-2}$. In other words, $(n-1)$-point single-gluon amplitudes can be extracted from the $n$-point scalar amplitude!

\paragraph{Flat space limit}
It is shown in~\cite{Alday:2021odx} that, with $\delta_{ij}=R^2s_{ij}$, the leading terms of $\mathcal M_n^{(s)}$ in the limit $R\to\infty$ matches the flat space color-ordered $n$-gluon amplitude, with $\epsilon_i\cdot p_j=0$ and $\epsilon_i\cdot\epsilon_j=\langle ij\rangle^2=-V_{ij}$. Equivalently, this is the flat-space amplitude of $\frac n2$ pairs of scalars in Yang-Mills-scalar theory~\cite{Cachazo:2013iea, Cachazo:2014xea}, which have been computed explicitly through $n=12$. For even $n$, everything is clear, and $\mathcal M_n^{(s)}\sim\delta^{2-\frac n2}$. For example, with $\epsilon\cdot p=0$,
\begin{equation}
    \mathcal A_4^\text{flat}=(\epsilon_1\cdot\epsilon_2)(\epsilon_3\cdot\epsilon_4)\frac{s_{12}+s_{23}}{s_{12}}+(1\leftrightarrow3)-(\epsilon_1\cdot\epsilon_3)(\epsilon_2\cdot\epsilon_4).
\end{equation}
Using $V_{13}V_{24}=V_{12}V_{34}+V_{14}V_{23}-2V_{1234}$ and writing $\mathcal X_{ij}$ in terms of $\delta_{ij}$, we can check that this matches the leading terms of the correct $n=4$ answer (up to overall normalization):
\begin{equation}\label{eq:m4}
    \begin{aligned}
        \mathcal M_4^{(s)}&=2 \left(\frac1{\mathcal X_{13}}+\frac1{\mathcal X_{24}}-1\right)V_{1234}\\
        &-\frac{2+\mathcal X_{24}}{\mathcal X_{13}}V_{12}V_{34}-\frac{2+\mathcal X_{13}}{\mathcal X_{24}}V_{14}V_{23}.
    \end{aligned}
\end{equation}
As an aside, it is a coincidence that the number $(n-1)!!$ of $(\epsilon\cdot\epsilon)^{\frac n2}$ terms equals $r_n$ for $n=4,6$. For $n\geq8$, these terms are not independent when translated to $V$. For odd $n$, the flat space amplitude vanishes due to the prescription $\epsilon_i\cdot p_j=0$. The power counting $s^{2-\frac n2}$ means that the order $\delta^{2-\lfloor\frac n2\rfloor}$ vanishes, and $\mathcal M_n^{(s)}\sim\delta^{2-\lceil\frac n2\rceil}$. A more careful argument using the formula proposed in~\cite{Penedones:2010ue} leads to the same conclusion.

\section{Constructing supergluon amplitudes}

It turns out that the properties and constraints satisfied by the Mellin amplitude discussed above are sufficient for a recursive construction of all tree-level supergluon amplitudes $\mathcal M_n^{(s)}$ for all $n$. Since $\mathcal M_{n-1}^{(v)}$ can be extracted from $\mathcal M_n^{(s)}$, we need only show that \emph{knowing $(\leq n-1)$-point scalar amplitudes and $(\leq n-2)$-point single-gluon amplitudes, we can construct the $n$-point scalar amplitude}.

The proof starts by noticing that $(\leq n-2)$-point scalar and single-gluon amplitudes completely fix the residue of $\mathcal M_n^{(s)}$ on all poles $\mathcal X_{ij}=-2m$ with $\|i-j\|\geq3$, where cyclic distance $\|i-j\|:=\min\{|i-j|,n-|i-j|\}$. Moreover, $(\leq n-1)$-point scalar amplitudes completely fix all incompatible channels. From these data, we can construct a rational function that can only differ from $\mathcal M_n^{(s)}$ by terms with only $\mathcal X_{i,i+2}=0$ poles~\footnote{$\mathcal X_{i,i+2}=-2m$ with $m>0$ never appears, because the shifted 3-point amplitudes $\mathcal M_{i,i+1,I}^{(s)(m)}$ and $\mathcal M_{i,i+1,I}^{(v)(m)}$ vanish for $m>0$.} and compatible traces. Then, we can write an ansatz for the possible difference, and completely fix it with constraints imposed by flat space limit and no-gluon kinematics.

Specifically, suppose $n=2n'+1$ is odd. Power counting $\mathcal M_n^{(s)}\sim\mathcal X^{1-n'}$, together with the fact that the ansatz only has $\mathcal X_{i,i+2}=0$ poles and compatible channels, implies that the ansatz consists of terms of the form
\begin{equation*}
    \frac{\text{constant}}{\mathcal X^{n'-1}}.
\end{equation*}
The constants are fixed by scalar exchanges at no-gluon kinematics because the polynomial remainder in \eqref{eq:NGfix} is ruled out by power counting.

Suppose $n=2n'$ is even. In the flat space limit, the leading terms are known, so the undetermined terms are subleading $\sim\mathcal X^{\leq1-n'}$. Since there are at most $n'$ simultaneous $\mathcal X_{i,i+2}$'s in the denominator, undetermined terms are of the form
\begin{equation*}
    \frac{\mathcal X^{\leq1}}{\mathcal X^{n'}}\quad\text{or }\quad\frac{\text{constant}}{\mathcal X^{n'-1}}.
\end{equation*}
For $n\geq6$, all such terms have no fewer than 2 simultaneous poles. To see that no-gluon kinematics is sufficient to fix the ansatz, simply note that we cannot construct a term $({\rm Numerator})/(\mathcal X_{ij}\mathcal X_{i'j'}\cdots)$ that vanishes at the no-gluon kinematics on every channel. For instance, for a term to vanish at no-gluon kinematics in both channels $\mathcal X_{13}=0$ and $\mathcal X_{35}=0$,
\begin{align*}
    {\rm Numerator}&=c_0\mathcal X_{13}+\sum_{i\neq1,2,3}c_i\left(\mathcal X_{2i}+1-\frac{\mathcal X_{1i}+\mathcal X_{3i}}2\right)\\
    &=d_0\mathcal X_{35}+\sum_{j\neq3,4,5}d_j\left(\mathcal X_{4j}+1-\frac{\mathcal X_{3j}+\mathcal X_{5j}}2\right).
\end{align*}
Comparing both expressions, we see that these force ${\rm Numerator}=0$.

Therefore, from $\mathcal M_3^{(s)}$ and $\mathcal M_4^{(s)}$ (which contain contact terms and cannot be fixed by factorization), we can recursively construct $\mathcal M_n^{(s)}$ for all $n$ as follows:
\begin{equation}\label{eq:sequence}
    \cdots\leadsto\mathcal M_{n}^{(s)}\leadsto\mathcal M_{n-1}^{(v)}\leadsto\mathcal M_{n+1}^{(s)}\leadsto\cdots
\end{equation}
It is satisfying to see that 3- and 4-point interactions determine the amplitudes of all $n$, much like flat-space Yang-Mills-scalar theory. As a by-product, we also obtain $\mathcal M_n^{(v)}$. We emphasize that this is a constructive procedure, which is quite efficient ($\leq5$ min to obtain $\mathcal M_8^{(s)}$).

\section{Discussion and outlook}
Based on a better organization of R-symmetry structures which leads to a clear separation of scalar and gluon exchanges, we have shown that all-$n$ supergluon tree amplitudes in AdS can be recursively constructed \eqref{eq:sequence}: we extract $(n{-}2)$-scalar-1-gluon amplitude from the $n$-scalar amplitude, which in turn determines the $(n{+}1)$-scalar amplitude. For instance, we could construct $\mathcal M_8^{(s)}$, knowing $\mathcal M_{\leq7}^{(s)}$ and (hence) $\mathcal M_{\leq6}^{(v)}$. In fact, we found in practice that even $\mathcal M_{\leq5}^{(v)}$ suffices! Another observation is that $\mathcal M_{1\cdots(k-1)I}^{(s)(m)}=0$ for $m\geq\lfloor\frac{k}2\rfloor$, and $\mathcal M_{1\cdots(k-1)I}^{(v)(m)}=0$ for $m\geq\lfloor\frac{k-1}2\rfloor$, which explains the truncation of poles $\mathcal X_{ij}=-2m$ at $m\leq\lfloor\frac{\|j-i\|}2\rfloor-1$ in any $\mathcal M_n^{(s)}$. We will discuss these matters in detail in a forthcoming paper~\cite{Cao:2024bky}.

Our results provide more data for studying color-kinematics duality and double copy in AdS~\cite{Zhou:2021gnu}. In addition, knowing the higher-point amplitudes, we can search for a set of Feynman rules. This will provide a better understanding of the bulk Lagrangian, as well as generalizing the Mellin space Feynman rules for scalars~\cite{Paulos:2011ie} and pure Yang-Mills~\cite{Chu:2023pea,Chu:2023kpe}.

Of course it would be highly desirable to apply similar methods to tree amplitudes with higher Kaluza-Klein modes ($\tau>2$), and eventually at loop level. We are also very interested in adopting this method for bootstrapping supergravity amplitudes in AdS, as a generalization of the beautiful $n=5$ results in~\cite{Goncalves:2019znr,Goncalves:2023oyx}. Note that the R-symmetry basis and flat-space results~\cite{Cachazo:2014xea} are available, and an immediate target would be the $n=6$ supergravity amplitude. 

We observe some universal behavior of our results, besides the ``scaffolding'' relation between a single gluon and a pair of scalars. For example, we find intriguing new structures such as ``leading singularities", {\it i.e.} maximal residues, which take a form that resemble flat-space result in $X$-variables. Our results and their generalizations strongly suggest that a possible combinatorial/geometric picture exists for AdS supergluon amplitudes, much like the scalar-scaffolding picture for gluons in flat space~\cite{Arkani-Hamed:2023jry}.

\section*{Acknowledgments}
It is our pleasure to thank Luis F. Alday and Xinan Zhou for inspiring discussions and for sharing their results together with Vasco Goncalves and Maria Nocchi on $n=6$ amplitudes~\cite{Alday:2023kfm}. We also thank Xiang Li for collaborations on related projects. This work has been supported by the National Natural Science Foundation of China under Grant No. 12225510, 11935013, 12047503, 12247103, and by the New Cornerstone Science Foundation through the XPLORER PRIZE.

\bibliographystyle{physics}
\bibliography{reference.bib}

\appendix
\widetext
\begin{center}
    \textbf{\large Supplemental Material}
\end{center}

\section{Details of Mellin factorization}~\label{sec:fac}
The Mellin amplitude $\mathcal M(\delta_{ij})$ of a correlator of scalar operators is defined by
\begin{equation}
\left\langle\mathcal{O}_1(x_1) \cdots \mathcal{O}_n(x_n)\right\rangle=\int[{\rm d}\delta]\mathcal M(\delta_{i j}) \prod_{1 \leq i<j \leq n} \frac{\Gamma(\delta_{ij})}{(x_{i j})^{2\delta_{i j}}}.
\end{equation}
Here, the Mellin variables satisfy constraints as if $\delta_{ij}=p_i\cdot p_j$ for auxiliary momenta with $\sum_ip_i=0$ and $p_i^2=-\Delta_i$, where $\Delta_i$ is the conformal dimension of $\mathcal O_i$:
\begin{equation}
\sum_{i=1}^n \delta_{i j}=0, \quad \delta_{i j}=\delta_{j i}, \quad \delta_{i i}=-\Delta_i.
\end{equation}
These constraints reduce the number of independent Mellin variables to $\frac12n(n-3)$. The integration is performed by integrating any set of independent Mellin variables parallel to the imaginary axis.

The poles of $\mathcal M(\delta_{ij})$ are determined by primary operators appearing in both $\prod_{a\in L}\mathcal O_a(x_a)$ and $\prod_{i\in R}\mathcal O_i(x_i)$ OPEs. Without loss of generality, let us consider the case where $L=\{1,\cdots,k-1\}$ and $R=\{k,\cdots n\}$. For each exchanged primary operator $\mathcal O_I$ with dimension $\Delta$ and spin $J$, $\mathcal M(\delta_{ij})$ has a tower of poles:
\begin{equation}
\mathcal M \asymp \frac{\mathcal{Q}_m}{\delta_{L R}-(\Delta-J+2 m)}, \quad m=0,1,2, \cdots
\end{equation}
where $m$ indicates the descendant level of the operator exchanged; \emph{e.g.}, $m=0$ corresponds to a primary exchange and $m=2$ corresponds to a level-2 descendant. In the denominator we have 
\begin{equation}
\delta_{L R}=-\left(\sum_{a=1}^{k-1} p_a\right)^2=\sum_{a=1}^{k-1} \sum_{i=k}^n \delta_{a i}=-{\cal X}_{1k}+2.
\end{equation}

Let us consider the case of scalar exchange ($J=0$) first. For a level-$m$ descendant,
\begin{equation}
    \mathcal Q_m=\frac{-2\Gamma(\Delta)m!}{(1+\Delta-\frac d2)_m}\mathcal M_L^{(m)}\mathcal M_R^{(m)},
\end{equation}
where the spacetime dimension $d$ is the CFT dimension of ${\rm AdS}_{d+1}/{\rm CFT}_d$, and the Pochhammer symbol $(x)_m:=\Gamma(x+m)/\Gamma(x)$. For $m=0$, we simply have $\mathcal M_L^{(0)}=\mathcal M_L$, the Mellin amplitude of $\langle\mathcal O_1(x_1)\cdots\mathcal O_{k-1}(x_{k-1})\mathcal O_I(x_I)\rangle$. In general,
\begin{equation}\label{eq:desShift}
    \mathcal M_L^{(m)}=\sum_{\mathclap{\substack{n_{ab}\geq0\\\sum n_{ab}=m}}}\mathcal M_L(\delta_{ab}+n_{ab})\ \ \prod_{\mathclap{1\leq a<b\leq k-1}}\ \ \frac{(\delta_{ab})_{n_{ab}}}{n_{ab}!}.
\end{equation}
The definition of $\mathcal M_R^{(m)}$ is similar.

Eq. \eqref{eq:desShift} deserves a few comments. The $k$-point amplitude $\mathcal M_L$ is usually presented as a function of $\frac12k(k-1)$ constrained Mellin variables. To use \eqref{eq:desShift}, we first need to solve $\delta_{aI}=-\sum_{b=1}^{k-1}\delta_{ab}$ using ``momentum conservation'' and write it in terms of $\frac12(k-1)(k-2)$ still constrained variables $\{\delta_{ab}\}_{1\leq a<b<k}$. The remaining constraint is $\delta_{LR}=\Delta+2m$, which reduces $\frac12(k-1)(k-2)$ to $\frac12k(k-3)$ independent variables. In terms of the redundant set of variables $\{\delta_{ab}\}_{1\leq a<b<k}$, $\mathcal M_L(\delta_{ab})$ does not have a unique functional form. However, the claim is that \eqref{eq:desShift} gives the same result for any such functional form $\mathcal M_L(\delta_{ab})$, provided we sum over all ways $\{n_{ab}\}$ of partitioning $m$ into $\frac12(k-1)(k-2)$ pieces. For example, using the notation in the main text,
\begin{equation}
    M^{(s)}(1234)=\frac2{\mathcal X_{13}}+\frac2{\mathcal X_{24}}-2=\underbrace{\frac1{\delta_{12}-1}+\frac1{\delta_{23}-1}-2}_{\mathcal A(\delta_{12},\delta_{13},\delta_{23})}=\underbrace{\frac1{1-\delta_{12}-\delta_{13}}+\frac1{1-\delta_{13}-\delta_{23}}-2}_{\mathcal B(\delta_{12},\delta_{13},\delta_{23})}.
\end{equation}
Note that we have solved $s_{a4}$ using ``momentum conservation'' as mentioned above. Plugging into \eqref{eq:desShift} for $m=1$,
\begin{align*}
    \mathcal A^{(1)}&=\delta_{12}\mathcal A(\delta_{12}+1,\delta_{13},\delta_{23})+\delta_{13}\overbrace{\mathcal A(\delta_{12},\delta_{13}+1,\delta_{23})}^{=\mathcal A(\delta_{12},\delta_{13},\delta_{23})}+\delta_{23}\mathcal A(\delta_{12},\delta_{13},\delta_{23}+1)\\
    &=\left(1+\frac{\delta_{12}}{\delta_{23}-1}-2\delta_{12}\right)+\left(\frac{\delta_{13}}{\delta_{12}-1}+\frac{\delta_{13}}{\delta_{23}-1}-2\delta_{13}\right)+\left(\frac{\delta_{23}}{\delta_{12}-1}+1-2\delta_{23}\right).
\end{align*}
Using the fact that $\delta_{LR}=6-2(\delta_{12}+\delta_{13}+\delta_{23})=\Delta+2m$ for the level-$m$ descendant, we have $\delta_{12}+\delta_{13}+\delta_{23}=1$. Without loss of generality, solve $\delta_{13}=1-\delta_{12}-\delta_{23}$ to arrive at $\mathcal A^{(1)}(\delta_{12},\delta_{23})=-2$. Similarly, for the functional form $\mathcal B(\delta_{12},\delta_{13},\delta_{23})$,
\begin{align*}
    \mathcal B^{(1)}&=\delta_{12}\mathcal B(\delta_{12}+1,\delta_{13},\delta_{23})+\delta_{13}\overbrace{\mathcal B(\delta_{12},\delta_{13}+1,\delta_{23})}^{\neq\mathcal B(\delta_{12},\delta_{13},\delta_{23})}+\delta_{23}\mathcal B(\delta_{12},\delta_{13},\delta_{23}+1)\\
    &=-\delta_{12}\left(\frac1{\delta_{12}+\delta_{13}}+\frac1{\delta_{13}+\delta_{23}-1}+2\right)+\cdots+\cdots.
\end{align*}
Once again, solving $\delta_{13}=1-\delta_{12}-\delta_{23}$ leads to $\mathcal B^{(1)}(\delta_{12},\delta_{23})=-2$.

Let us now turn to the exchange of spinning operators. For the purpose of this paper, we only consider vector exchanges ($J=1$). In order to specify $\mathcal Q_m$, we first need to consider the Mellin representation of single-vector correlators, \emph{i.e.}, the left and right half amplitudes. Mellin amplitudes of spinning correlators are most convenient to define using the embedding space formalism where the action of the conformal group is linearized. Each point $x^\mu \in \mathbb{R}^d$ is lifted to a null ray $P^A \in \mathbb{R}^{1, d+1}$ which satisfies $P\cdot P=0$ and $P\sim \lambda P$. Operators of dimension $\Delta$, spin $J$ are homogeneous functions of $P$ and $Z$:
\begin{equation}
    \mathcal{O}(\lambda P+\alpha Z)=\lambda^{-\Delta} \alpha^J \mathcal{O}(P, Z),
\end{equation}
where the polarization $Z^A \in \mathbb{R}^{1, d+1}$ encodes the tensor structure and satisfies $Z\cdot Z=P \cdot Z=0$. We further impose the transversality condition
\begin{equation}\label{eq:trans}
  \mathcal{O}(P, Z+\beta P)=\mathcal{O}(P, Z).
\end{equation}

The Mellin amplitude with $(k-1)$ scalars and 1 vector is defined by
\begin{equation}
\left\langle\mathcal{O}(P, Z) \mathcal{O}_1\left(P_1\right) \ldots \mathcal{O}_{k-1}\left(P_{k-1}\right)\right\rangle=\sum_{c=1}^{k-1}\left(Z \cdot P_c\right) \int[d \delta] \mathcal{M}^c 
\times \prod_{1\leq a<b<k} \frac{\Gamma\left(\delta_{ab}\right)}{\left(-2 P_a \cdot P_b\right)^{\delta_{ab}}} \prod_{a=1}^{k-1} \frac{\Gamma\left(\delta_{0 a}+\delta_a^c\right)}{\left(-2 P_a \cdot P_0\right)^{\delta_{0 a}+\mathbf{\delta}_a^c}},
\end{equation}
where $\delta_a^c$ is the Kronecker delta, and the Mellin variables satisfy
\begin{equation}
\delta_{0a}=-\sum_{b=1}^{k-1} \delta_{ab}, \quad \delta_{aa}=-\Delta_a, \quad \sum_{a=1}^{k-1} \delta_{0a}=\Delta-1.
\end{equation}
From \eqref{eq:trans}, we obtain the ``gauge invariance'' condition: 
\begin{equation}
\sum_{a=1}^{k-1} \delta_{0a} M^a=0.
\end{equation}

Finally, let us present the expression $\mathcal Q_m$ due to vector exchange, in the special case where the exchanged operator is a conserved current which has $\Delta=d-1$:
\begin{equation}
    \mathcal Q_m=\Delta\Gamma(\Delta-1)\frac{m!}{(\frac d2)_m}\sum_{a=1}^{k-1}\sum_{i=k}^n\delta_{ai}\mathcal M_L^{a(m)}\mathcal M_R^{i(m)}.
\end{equation}
Here, the shift prescription is exactly the same as \eqref{eq:desShift}.

\section{Explicit results for amplitudes up to $n=6$}~\label{sec:result}
We record the complete supergluon amplitudes for $n\leq8$ and single-gluon amplitudes for $n\leq6$ in an ancillary file. Here we present compact expressions for $n\leq 6$ supergluon amplitudes as well as new results for $n=5$ spinning amplitudes. 

Explicit results for $n=5$ supergluon amplitude are:
\begin{align}
    M^{(s)}(12345)&=4 \left(\frac{-1}{\mathcal X_{13}}+\frac{1}{\mathcal X_{13}\mathcal X_{14}}\right)+4\text{ cyclic},\label{5pt_single}\\
    M^{(s)}(12;345)&=\frac{4}{\mathcal X_{13}}-\frac1{\mathcal X_{13}}\left(\frac{2(\mathcal X_{24}+2)}{\mathcal X_{14}}+\frac{2(\mathcal X_{25}+2)}{\mathcal X_{35}}\right).
\end{align}

Explicit results for $n=6$ supergluon amplitude are:
\begin{align}
    \frac{M^{(s)}(123456)}{8}&=\left[\left(\frac 1 {{\cal X}_{14}}+\frac{1}{{\cal X}_{14}+2}\right)+ 2~{\rm cyclic}\right]\nonumber\\
    &-\left[\left(\frac 1{{\cal X}_{13}{\cal X}_{14}} + \frac 1 {{\cal X}_{13} {\cal X}_{15}} + \frac 1 {{\cal X}_{14} {\cal X}_{15}}\right)+ 5~{\rm cyclic} + \frac 1 {{\cal X}_{13} {\cal X}_{46}} + 2~ {\rm cyclic}\right]\nonumber \\
    &+\left[\left(\frac 1 {{\cal X}_{13}{\cal X}_{14} {\cal X}_{15}}+ \frac 1 {{\cal X}_{13} {\cal X}_{14} {\cal X}_{46}}\right) + 5~{\rm cyclic} + \frac 1 {{\cal X}_{13} {\cal X}_{15} {\cal X}_{35}} + 1~{\rm cyclic}\right],\label{6pt-single}\\
    \frac{M^{(s)}(12;3456)}{4}&=\frac{1}{{\cal X}_{13}}\left(-2+\frac{2}{{\cal X}_{15}}+ \frac{2}{{\cal X}_{35}}+\frac{2}{{\cal X}_{46}}-\frac{{\cal X}_{25}+2}{{\cal X}_{15} {\cal X}_{35}}+({\cal X}_{24}+2) \left(\frac{1}{{\cal X}_{14}}+\frac{1}{{\cal X}_{14}+2}-\frac{1}{{\cal X}_{14}{\cal X}_{15}}-\frac{1}{{\cal X}_{14} {\cal X}_{46}}\right)\right.\nonumber \\
    &\quad\quad\quad\quad\left.+\,({\cal X}_{26}+2) \left(\frac{1}{{\cal X}_{36}}+\frac{1}{{\cal X}_{36}+2}-\frac{1}{{\cal X}_{36}{\cal X}_{46}}-\frac{1}{{\cal X}_{35} {\cal X}_{36}}\right)\right),\\
    \frac{M^{(s)}(123;456)}{4}&=\frac{1}{{\cal X}_{14}}\left(\frac{2}{{\cal X}_{13}}+ \frac{2}{{\cal X}_{15}}+\frac{2}{{\cal X}_{24}}+\frac{2}{{\cal X}_{46}}-\frac{{\cal X}_{25}+2}{{\cal X}_{15}{\cal X}_{24}}-\frac{{\cal X}_{35}+2}{{\cal X}_{13}{\cal X}_{15}} -\frac{{\cal X}_{26}+2}{{\cal X}_{24}{\cal X}_{46}} -\frac{{\cal X}_{36}+2}{{\cal X}_{13}{\cal X}_{46}} \right)\nonumber\\
    &-\left(\frac{1}{{\cal X}_{14}}+\frac{1}{{\cal X}_{14}+2}\right),\\
    \frac{M^{(s)}(12;36;45)}{2}&=\frac{1}{{\cal X}_{13} {\cal X}_{46}}\left(-2 {\cal X}_{25}-8+\frac{\left({\cal X}_{15}+2\right) \left({\cal X}_{24}+2\right)}{{\cal X}_{14}}+\frac{\left({\cal X}_{15}+2\right) \left({\cal X}_{24}+2\right)}{{\cal X}_{14}+2}\right.\nonumber\\
    &\quad\quad\quad\quad\quad\ \left.+\,\frac{\left({\cal X}_{26}+2\right) \left({\cal X}_{35}+2\right)}{{\cal X}_{36}}+\frac{\left({\cal X}_{26}+2\right) \left({\cal X}_{35}+2\right)}{{\cal X}_{36}+2}\right),\\
    \frac{M^{(s)}(12;34;56)}{2}&=\left[\frac{1}{{\cal X}_{13} {\cal X}_{15}}\left(2\left({\cal X}_{14}-{\cal X}_{24}+{\cal X}_{26}-{\cal X}_{46}-2\right)+\frac{\left({\cal X}_{24}+2\right) \left({\cal X}_{46}+2\right)}{ {\cal X}_{14}}+\frac{\left({\cal X}_{24}+2\right) \left({\cal X}_{46}+2\right)}{ {\cal X}_{14}+2}\right)+2\text{cyclic}\right]\nonumber\\
    &+\frac{2}{{\cal X}_{13} {\cal X}_{15} {\cal X}_{35}}\left(4+{\cal X}_{46}+{\cal X}_{24}+{\cal X}_{26}+{\cal X}_{36} {\cal X}_{24}-{\cal X}_{14} {\cal X}_{25}+{\cal X}_{14} {\cal X}_{26}-{\cal X}_{14} {\cal X}_{36}-{\cal X}_{25} {\cal X}_{36}+{\cal X}_{25} {\cal X}_{46}\right).
\end{align}
Here we pause to discuss some nice structures already seen for single-trace amplitudes. As we have mentioned, any single-trace amplitude is given by the flat-space amplitude of a theory with ${\rm tr}(\phi^3-\phi^4)$ interactions, except for terms with descendant poles.  For $M^{(s)}(1234)$, we have $\frac{1}{{\cal X}_{13}}+ \frac{1}{{\cal X}_{24}}$ (cubic diagrams) minus $1$ (quartic diagram), or in the language of associahedron~\cite{Arkani-Hamed:2017mur}, the 2 vertices and 1 edge of a line interval. Similarly, for $M^{(s)}(12345)$ in \eqref{5pt_single}, we have $5$ terms of the form $\frac 1 {{\cal X}_{13} {\cal X}_{14}}$ (vertices) minus $5$ of the form $\frac 1 {{\cal X}_{13}}$ (edges). For $n=6$ in \eqref{6pt-single}, there are $14$ terms corresponding to vertices (with $+1$ coefficient, third line), $21$ terms corresponding to edges (with $-1$ coefficient, second line), and $3$ ``square" faces (with $+1$ coefficient, first line). There are no terms corresponding to ``pentagonal'' or ``hexagonal'' faces or the bulk itself due to absence of 5-point  or 6-point vertices. 

For $n$-point single-trace amplitude we write
$M^{(s)}(12\cdots n)=c(A^{(0)}_n+ R_n)$ where $c$ is an overall constant, and $A^{(0)}_n$ contains all terms with only primary poles given by ${\rm tr}(\phi^3-\phi^4)$ amplitude ({\it e.g.} there are $154$ terms for $n=7$ and $654$ terms for $n=8$, with alternating signs as above); the remainder $R_n$ contains descendant poles only: $R_n=0$ for $n=4,5$ but becomes non-trivial for $n\geq 6$. For example, we have exactly $3$ terms for the remainder of $n=6$,
\begin{equation}
    R_6=\frac{1}{{\cal X}_{1 4}+2}+\frac{1}{{\cal X}_{2 5}+2}+\frac{1}{{\cal X}_{3 6}+2},
\end{equation}
and $4\times 7$ terms in $R_7$: 
\begin{equation}
R_7=\left(\frac{1}{{\cal X}_{13}({\cal X}_{15}+2)}+\frac{1}{{\cal X}_{24}({\cal X}_{15}+2)}+\frac{1}{{\cal X}_{35}({\cal X}_{15}+2)}+\frac{1}{({\cal X}_{1 4}+2)({\cal X}_{1 5}+2)}\right)+6~{\rm cyclic}.
\end{equation}
There is no difficulty to obtain an all-$n$ formula for $R_n$ since $M^{(s)}(12\cdots n)$ can be constructed purely from scalar factorizations. We hope to obtain all-$n$ results for double- and triple-trace amplitudes, which do not require explicit form of spinning amplitudes. 

From the $n=6$ results above, one can immediately extract $n=5$ spinning amplitudes by imposing gauge invariance:
\begin{align}
M_5^{(v) 1}&=2 V_{1234} \left(\frac{1}{{\cal X}_{13}{\cal X}_{14}} + \frac{1}{{\cal X}_{14}{\cal X}_{24}} +\frac{1}{{\cal X}_{24}{\cal X}_{25}} +\frac{1}{{\cal X}_{25}{\cal X}_{35}} +\frac{1}{{\cal X}_{13}{\cal X}_{35}} -\frac{1}{{\cal X}_{14}}-\frac{1}{{\cal X}_{25}}-\frac{2}{{\cal X}_{25}+2}\right)\nonumber\\
&+ V_{14} V_{23} \frac{1}{{\cal X}_{24}}  \left( 2-\frac{2+{\cal X}_{13}}{{\cal X}_{14}} -\frac{2+{\cal X}_{35}}{{\cal X}_{25}} - \frac{2(2+{\cal X}_{35})}{{\cal X}_{25}+2} \right)\nonumber\\
&+ V_{12} V_{34} \left(\frac{-1}{{\cal X}_{13}} (2+ \frac{2+{\cal X}_{24}}{{\cal X}_{14}}) +\frac{1}{{\cal X}_{35}}(2+ \frac{2+{\cal X}_{24}}{{\cal X}_{25}} - \frac{2(2+{\cal X}_{24})}{{\cal X}_{25}+2})-\frac{4+{\cal X}_{14}+2 {\cal X}_{24}-{\cal X}_{25}}{{\cal X}_{13} {\cal X}_{35}}\right)\,,\\
M_5^{(v) 2}&=2 V_{1234} \left(\frac{1}{{\cal X}_{13}{\cal X}_{14}} + \frac{1}{{\cal X}_{14}{\cal X}_{24}} +\frac{1}{{\cal X}_{25}{\cal X}_{35}}-\frac{1}{{\cal X}_{24}{\cal X}_{25}}-\frac{1}{{\cal X}_{13}{\cal X}_{35}} -\frac{1}{{\cal X}_{14}}+\frac{1}{{\cal X}_{25}}\right)\nonumber\\
&+ V_{14} V_{23} \frac{1}{{\cal X}_{24}}  \left( 2-\frac{2+{\cal X}_{13}}{{\cal X}_{14}} -\frac{2+{\cal X}_{35}}{{\cal X}_{25}}\right)\nonumber\\
&+ V_{12} V_{34} \left(\frac{1}{{\cal X}_{13}} (2+ \frac{2+{\cal X}_{24}}{{\cal X}_{14}}) - \frac{1}{{\cal X}_{35}}(2+ \frac{2+{\cal X}_{24}}{{\cal X}_{25}})+\frac{3{\cal X}_{14}-2 {\cal X}_{24}+{\cal X}_{25}}{{\cal X}_{13} {\cal X}_{35}}\right)\,,
\end{align}
and $M_5^{(v) 4}, M_5^{(v) 3}$ are related by reflection symmetry. 

\paragraph{Leading singularities} We present some examples of leading singularities obtained by taking $n{-}3$ residues (for primary poles) with compatible ${\cal X}_\alpha=0$ (for any triangulation with $n{-}3$ chords $\alpha$ of the $n$-gon):
\begin{equation}
{\cal L}(\{{\cal X_\alpha}\}):=\mathop{\rm Res}_{\{{\cal X}_\alpha=0\}} {\cal M}_{12 \cdots n}\,.  
\end{equation}
These are polynomials of ${\cal X}$ variables (dressed with certain R-structures), which can be obtained by gluing $3$-point building blocks together. They take particularly suggestive forms resembling the flat-space leading singularities in $X$ variables~\cite{Arkani-Hamed:2023jry}. For example, for $n=5$, 
\begin{equation}
{\cal L}({\cal X}_{13}, {\cal X}_{14})=2 V_{12345}- V_{12} V_{34} (2+{\cal X}_{24})- V_{123} V_{45} (2+{\cal X}_{35})\,,  
\end{equation}
and for $n=6$ with different triangulations:
\begin{align}
    {\cal L}({\cal X}_{13}, {\cal X}_{14}, {\cal X}_{15})&=4 V_{123456}-2\left( V_{123} V_{456}(2+{\cal X}_{35})+V_{12} V_{3456} (2+{\cal X}_{24})+V_{1234} V_{56}(2+{\cal X}_{46})\right)\nonumber\\
    &+V_{12} V_{34} V_{56}(2+{\cal X}_{24})(2+{\cal X}_{46}),\\
    {\cal L}({\cal X}_{13}, {\cal X}_{35}, {\cal X}_{15})&= 4 V_{123456}-2 \left(V_{12} V_{3456} (2+{\cal X}_{25})+V_{34} V_{1256} (2+ {\cal X}_{14})+V_{56} V_{1234} (2+{\cal X}_{36}) \right)\nonumber\\
    &+ 2 V_{12} V_{34} V_{56}\left(4+ {\cal X}_{24}+{\cal X}_{46}+ {\cal X}_{26}-{\cal X}_{14}({\cal X}_{25}-{\cal X}_{26})-{\cal X}_{25}({\cal X}_{36}-{\cal X}_{46})-{\cal X}_{36}({\cal X}_{14}-{\cal X}_{24})\right)\,.
\end{align}
The latter gives the scalar-scaffolded $3$-gluon amplitudes in Yang-Mills theory~\cite{Arkani-Hamed:2023jry} in the flat-space limit.

\end{document}